# Quantum measurement: a game between observer and nature?

Lizhi Xin[1] and Houwen Xin[2, *]

[1]Building 59, 96 Jinzhai Road, Hefei, Anhui, P. R. China
[2]Department of Chemical physics USTC, Hefei, Anhui, P. R. China
[*]hxin@ustc.edu.cn

## ABSTRACT

What is the observer's role in quantum measurement? Obviously, observers prepare the apparatus, observe and interpret the measured results. Although the observer will have a certain influence on the measurement results by setting up the measuring apparatus, we don't believe human consciousness cause reducing of wave packet; also observers are certainly required to interpret the measured results with physical meanings. We believe observers build up their experience of the external world by playing games with nature, and then "decode" the nature based on their experiences. We propose a quantum decision theory approach to explain the role of the observer in quantum measurements, and pointed out that a set of quantum decision trees (strategies to answer natural questions with yes/no logic) can be optimized to deal with the challenges of nature through quantum genetic programming based on maximization of the expected value of the observers; Quantum decision trees can discover the dynamics rules of quantum entities and Just as classical mechanics uses the principle of least action to obtain the trajectories of particles, we use the principle of maximum expected value to approximately obtain the kinematic "trajectories" of quantum entities by learning from natural historical "events" (measured results); even we can "reconstruct" the past of quantum entity, because we don't know the prior information of quantum entity, it is very difficult to predict the future of the nature.

## Introduction

In physics, measurement is an act to collect numerical data that describes a property of an object. Usually, a measurement is made by comparing a quantity of an object with a standard unit so that a measuring device can obtain a value that matches the property of the measured object; in other words, measurement is an act to deliver a value by correlating the state of observed object with the pointer state of apparatus. Usually apparatus are set up by observer, and when the measurement is done, the measured results need to be observed and interpreted by an observer. The process of the measurement is shown as in Figure 1.

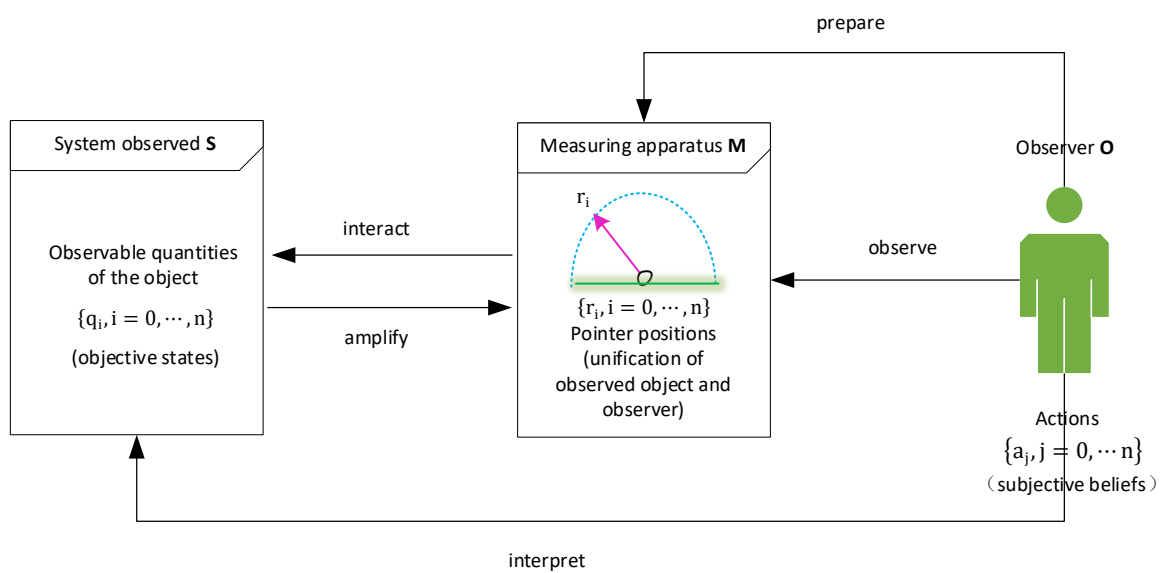

**Figure 1 Process of the measurement.**

A measurement has three parts:

- System observed **S** → $q_i = \{q_1, \cdots q_i, \cdots, q_n\}$ observable quantities (states) to be measured.

- Measuring device **M** → $r_i = \{r_1, \cdots r_i, \cdots, r_n\}$ pointer's states to be correlated $\left(1 \overset{to}{\to} 1\right)$ with the states of object $q_i = \{q_1, \cdots q_i, \cdots, q_n\}$ by interacting with the system observed and delivers (amplify) a value $r_j$ (pointer position).

- Observer **O** → prepare the apparatus, observe the measured results $r_i = \{r_1, \cdots r_i, \cdots, r_n\}$, and interpret the observed results with physical meanings. If observer doesn't "look at" the apparatus, he or she can subjectively take an action $a_i = \{a_1, \cdots a_i, \cdots, a_n\}$ to "compute" the "theoretical" results. In other words, observer "gambles" with nature to see if the "guessing" results are right (compared with the measured results), so that the observer can learn about nature and build-up his or her own experience of nature.

The process of a measurement is as follows:

1) Prepare($t = t_0$): the pointer state of the apparatus is set to initial state by an observer.

$$|\psi^M(t_0)\rangle = |r_0\rangle \qquad (1).$$

2) Measure($t_0 < t < t_1$): apparatus interacts with the measured system and amplify the signal.

   a) Classical measurement:

   $$|r_0\rangle \otimes |q_i\rangle \to |r_i\rangle|q_i\rangle \qquad (2a)$$

   b) Quantum measurement:

   $$|\psi^{S+M}(t_0)\rangle = |r_0\rangle \otimes \left(\sum_{i=1}^n c_i |q_i\rangle\right) \to \sum_{i=1}^n c_i |r_i\rangle|q_i\rangle = |\psi^{S+M}(t_0 < t < t_1)\rangle \qquad (2b)$$

3) Observe($t = t_1$): an observer observes and interprets the measured results.

$$|\psi^M(t_1)\rangle = |r_i\rangle, i = 1, \cdots, i, \cdots n. \qquad (3)$$

For the observer, the final measurement result is a definite pointer state $|r_i\rangle$ for both classical and quantum measurement. The difference is that the results of the classical measurements are consistent with the properties of the observed system (except for a negligible error), while the results of the quantum measurements are inconsistent with the properties of the observed system (the pointer state $r_i$ for each measurement is indeterminate and can only occur in Bonn probabilities $\omega_i = |c_i|^2$.

- Classical measurement postulate: the objective properties of the observed system are independent of the measurement and are not interfered with by the apparatus. The observed system has a definite state $q_i$, which corresponds to the pointer state $r_i$ with a definite one-to-one mapping, and it can be verified by the system measurement. After the measurement, the observed system's $q_i$ can be inferred from the result $r_i$. The classical theory does not have interpretation problem, because the measured results are consistent with the predictions of the classical theory. There is no problem at all with the classical measurements from 2) to 3).

- Quantum measurement postulate: there are multiple possible states $\{q_1, \cdots q_i, \cdots, q_n\}$ of the system observed, and prior information regarding the system is incomplete. Because the interaction of the observed system and the apparatus "disturbs" with each other's state, the pointer state $r_i$ of an apparatus can only be pointed to $q_i$ with a certain probability $\omega_i = |c_i|^2$ (Born rule), so observer cannot accurately infer which state $q_i \in \{q_1, \cdots q_i, \cdots, q_n\}$ the system was "in" before the measurement. The core of the quantum measurement problem is[1-7]: how does it happen from 2) to 3)?

Base on the above analysis, there are three problems in quantum measurement:

1) S Reality problem: does state (wave function) describe the reality of quantum entity or just a mathematical tool? Is wave function a complete description of physical reality?

2) M Entangle problem: how pointer states $\{r_1, \cdots r_i, \cdots, r_n\}$ "entangle" the observed system states $\{q_1, \cdots q_i, \cdots, q_n\}$ with the define $1 \xrightarrow{to} 1$ mapping? Where is the sharp boundary between the quantum world and the classical world?

3) O Interpretation problem: for a single measurement there is no so-called objective probability of repeated measurements; and in this case for the observer, how to interpret the inherent uncertain state of quantum systems? Does human consciousness cause reducing of wave packet? Or does the measurement process not require the existence of an observer?

We don't believe human consciousness cause reducing of wave packet; also we believe an observer is required to interpret the measured results. We propose a quantum decision approach for quantum measurement. For an observer to measure the properties of a quantum entity is like to play a game with the nature: nature makes his "choice", and the observer "bet" on it. In other words, an observer has to make a decision under uncertainty with incomplete information regarding nature's "choice"; through this learning progress, the observer gradually built-up his (her) own experience of the nature in his (her) memory for future decision-making. Figure 2 shows a modified Schrödinger's cat thought experiment for an observer to play game with nature (quantum system).

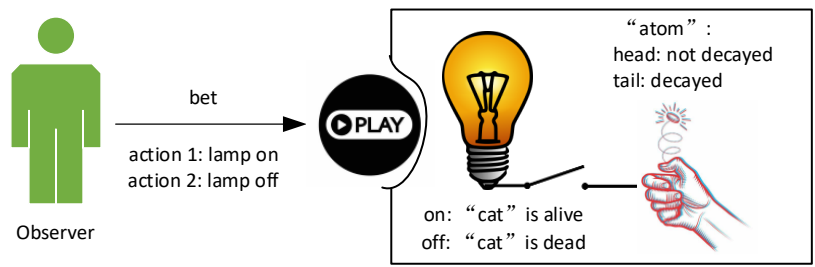

Figure 2 Modified Schrödinger's cat thought experiment.

The rules of the game are as follows:

1) A digital "coin" will be thrown within an hour, if head is up (0), nothing happens, the lamp is still on; if the tail is up (1), then a switch will be closed to cut the lamp off

2) If the lamp is on and an observer bets the lamp is on, the observer win the game, otherwise the observer lose it; if the lamp is off and an observer bets the lamp is off, the observer win the game, otherwise the observer lose it.

The modified Schrödinger's cat thought experiment can be simulated by a computer program which randomly generates N (=10000) results. We can define a random variable x to represent the fluctuation of the digital cat's state as in (4). The generated data series $\{(q_1, x_1), \cdots, (q_k, x_k), (q_{k+1}, x_{k+1}), \cdots, (q_N, x_N)\}$ is shown as in Table 1; $q_k$ denotes the state of "atom": 0 (not decayed) and 1 (decayed); if the "atom" has not decayed then variable x increases by 1, else variable x decreases by 1. The uncertainty of the cat's state is represented by the volatility of variable x as shown in Figure 3.

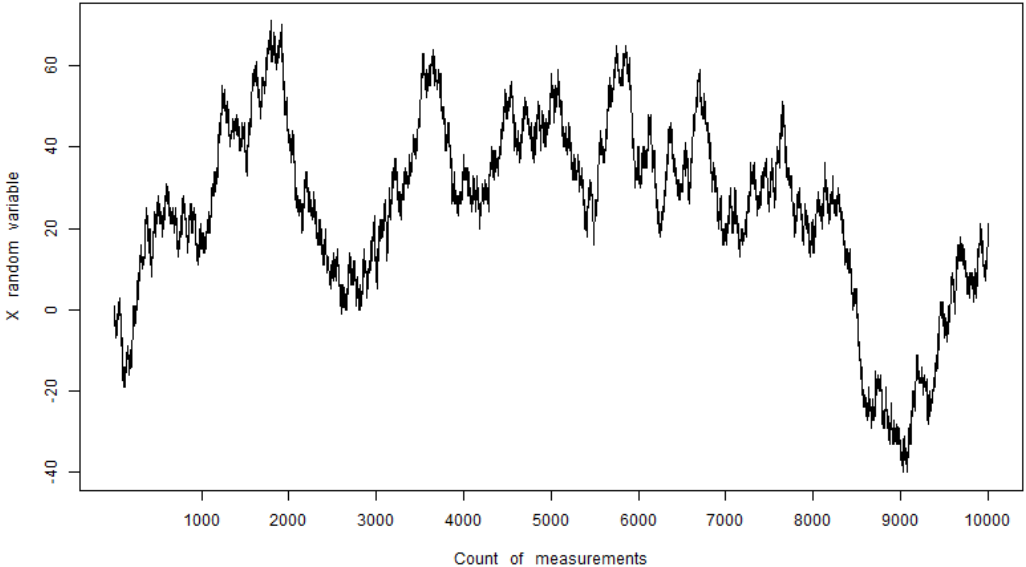

**Figure 3 Generated data series (10000 results) simulated by a computer program for the modified Schrödinger's cat thought experiment.**

$$x_{k=0} = 0; \tag{4a}$$

$$x_{k+1} = x_k + \begin{cases} -1, & \text{if "decayed"} \\ 1, & \text{if not "decayed"} \end{cases} \quad k = 1, \cdots, N \tag{4b}$$

| number | state ($q_k$) | value ($x_k$) |
|---|---|---|
| 1 | 0 | 1 |
| 2 | 1 | 0 |
| 3 | 1 | -1 |
| ⋮ | ⋮ | ⋮ |
| ⋮ | ⋮ | ⋮ |
| 5000 | 0 | 56 |
| ⋮ | ⋮ | ⋮ |
| ⋮ | ⋮ | ⋮ |
| 9998 | 0 | 20 |
| 9999 | 0 | 21 |
| N=10000 | 1 | 20 |

**Table 1 10000 generated results of digital "cat" experiment.**

Now if we ask this question: is it possible that the observer can "beat" the nature (quantum system) in the game? In other words, is it possible that the observer can have a reasonable expectation of the natural state by learning historical results of repeated measurements on copies of the same system ("atom" + "cat")?

For a large amount of repeated measurements, there is an objective frequency (50% the lamp is on); but for a single game there is no so-called objective probability, the lamp is either on or off., and the observer either win or loss; in this case, only the observer has subjective degrees of belief that the lamp is on or off before the box is opened.

Because of nature's inherent uncertainty (atom decayed or not), we can only get answers ("cat" is alive or dead) by asking questions and play a game with nature. Before opening the "black box" to have a right answer, the observer can "compute" what the answer is, and can improve the successful rate of having the right answer by learning historical data based on the maximization of expected value. We propose a quantum expected value decision theory[8-9] for observers, and quantum genetic programming is applied to evolve "satisfactory" strategies for observers to "guess" (with degrees of belief) the natural state as best possible based on quantum expected value.

**Quantum expected value (qEV)**

| State / Action | $q_1$ | $q_2$ |
|---|---|---|
| $a_1$ | $r_{11}$ | $r_{12}$ |
| $a_2$ | $r_{21}$ | $r_{22}$ |

**Table 2** A pay-off table of a game between an observer and the quantum system observed (nature).

An observer subjectively chooses an action $a_i \in \{a_1, a_2\}$ where the atom's objective state is in $\varphi_j \in \{q_1, q_2\}$ when a bet is made, and the result of bets depends on both the state of the atom and choice of observer's brain shown in table 2. The state of the atom describes the objective world; it can be represented by the superposition of all possible states in terms of the Hilbert state space[10-12] as in (5).

$$|\psi\rangle_{atom} = c_1|q_1\rangle + c_2|q_2\rangle \quad |c_1|^2 + |c_2|^2 = 1 \quad (5)$$

Where $|q_1\rangle$ denotes a state in which the atom has not decayed and $|q_2\rangle$ denotes a state in which the atom has decayed. $|c_1|^2$ is the objective frequency of the decayed atom; $|c_2|^2$ is the objective frequency of the atom without decay.

The observer's mental state describes the subjective world; we postulate that when the observer is undecided in making a bet, it can be represented by superposition of all possible actions as in (6).

$$|\phi\rangle_{mental} = \mu_1|a_1\rangle + \mu_2|a_2\rangle \quad |\mu_1|^2 + |\mu_2|^2 = 1 \quad (6)$$

Where $|a_1\rangle$ denotes the observer's action to believe the "cat" is alive and $|a_2\rangle$ denotes the observer's action to believe the "cat" is dead. $|\mu_1|^2$ is observer's subjective probability (degrees of belief) in betting "cat" is alive; $|\mu_2|^2$ is observer's subjective probability in betting "cat" is dead.

The prior information of the quantum world is incomplete; the result of the observer's decision is uncertain and it can be denoted by a mixed state's density operator as a value operator in (7). Value operator is a sum of projection operators which projects the observer's degrees of belief onto an action of choice.

$$\hat{V} = p_1|a_1\rangle\langle a_1| + p_2|a_2\rangle\langle a_2| \quad p_1 + p_2 = 1 \quad (7)$$

The observer's mental state is transformed from a pure state into a mixed state, and then one of available actions is selected by the brain (choose an action $a_i$ with probability $p_i$) as in (8). Based on the information that the brain has before a bet is made, there is only the possibility of selecting an action with a subjective probability (degrees of belief), and it is the decision-making process that makes the potential possibility a reality. The decision process of an observer can be simulated by the continuous evolution of the value operator according to the environment (information). Information is the essence of people's subjective beliefs just like energy is the essence of the objective world. Valuable information can reduce uncertainty.

$$D: \hat{\rho} = |\phi\rangle\langle\phi| \rightarrow \hat{V} = p_1|a_1\rangle\langle a_1| + p_2|a_2\rangle\langle a_2| \xrightarrow{decision} |a_i\rangle\langle a_i|, i = 1,2 \quad (8)$$

Quantum expected value qEV can be represented as in (9).

$$\begin{aligned}
qEV &= \langle\psi|\hat{V}|\psi\rangle = (c_1^*\langle q_1| + c_2^*\langle q_2|)(p_1|a_1\rangle\langle a_1| + p_2|a_2\rangle\langle a_2|)(c_1|q_1\rangle + c_2|q_2\rangle) \\
&= p_1\omega_1|\langle a_1|q_1\rangle|^2 + p_1\omega_2|\langle a_1|q_2\rangle|^2 + p_2\omega_1|\langle a_2|q_1\rangle|^2 + p_2\omega_2|\langle a_2|q_2\rangle|^2 \\
&= p_1\omega_1 r_{11} + p_1\omega_2 r_{12} + p_2\omega_1 r_{21} + p_2\omega_2 r_{22} \\
&= \sum_{i=1,2} p_i \sum_{j=1,2} \omega_j r_{ij}
\end{aligned} \tag{9}$$

Where $p_i = |\mu_i|^2$ is a observer's subjective probability in choosing an action $a_i$, subjective probability represents the observer's degrees of belief in a single event; $\omega_j = |c_j|^2$ is the objective frequency at which state of the atom is in $q_j$, objective frequency represents the statistical results of multiple occurrences of objective states; matrix $r_{ij} = |\langle a_i|q_j\rangle|^2$ is the value when the decision was made, in which the observer choose an action $a_i$ where atom's state is in $q_j$ as in (10). The different actions that the observers took lead to different value; in other words; the value is "created" based on both observers' subjective beliefs and objective natural states.

$$r_{ij} = |\langle a_i|q_j\rangle|^2 = \begin{cases} 1, & i = j \\ -1, & i \neq j \end{cases} \tag{10}$$

## Quantum decision tree (qDT)

The value operator is a 2x2 matrix, and the value operator needs to be diagonalized first and then normalized to get probability $p_1$ and $p_2$ as in (11).

$$\hat{V} = \begin{bmatrix} V_{11} & V_{12} \\ V_{21} & V_{22} \end{bmatrix} \xrightarrow{\text{diagonalization}} \begin{bmatrix} \lambda_1 & 0 \\ 0 & \lambda_2 \end{bmatrix} \xrightarrow{\text{normalization}} \begin{bmatrix} p_1 & 0 \\ 0 & p_2 \end{bmatrix} = p_1|a_1\rangle\langle a_1| + p_2|a_2\rangle\langle a_2| \tag{11a}$$

$$|a_1\rangle = \begin{bmatrix}1\\0\end{bmatrix}, |a_2\rangle = \begin{bmatrix}0\\1\end{bmatrix};\ |a_1\rangle\langle a_1| = \begin{bmatrix}1&0\\0&0\end{bmatrix}, |a_2\rangle\langle a_2| = \begin{bmatrix}0&0\\0&1\end{bmatrix} \tag{11b}$$

A value operator $\hat{V}$, as a qDT, can be constructed from basic quantum gates[13-14] with logic operations. The qDT composes of different nodes and branches. There are two types of nodes, non-leaf nodes and leaf nodes. The non-leaf nodes are composed of the operation set F as in (12); the leaf nodes are composed of the data set T (quantum gates) as in (13). The construction process of a qDT is to randomly select a logic symbol from the operation set F as the root of the qDT, and then grows corresponding branches according to the nature of the operation symbol and so on until a leaf node is reached.

$$F = \{+(\text{ADD}),\quad *(\text{MULTIPLY}),\quad //(\text{OR})\} \tag{12}$$

$$T = \{H, X, Y, Z, S, D, T, I\} \tag{13a}$$

$$\begin{cases} H = \frac{1}{\sqrt{2}}\begin{bmatrix}1&1\\1&-1\end{bmatrix}\ X = \begin{bmatrix}0&1\\1&0\end{bmatrix}\ Y = \begin{bmatrix}0&-i\\i&0\end{bmatrix}\ Z = \begin{bmatrix}1&0\\0&-1\end{bmatrix} \\ S = \begin{bmatrix}1&0\\0&i\end{bmatrix}\ D = \begin{bmatrix}0&1\\-1&0\end{bmatrix}\ T = \begin{bmatrix}1&0\\0&e^{i\pi/4}\end{bmatrix}\ I = \begin{bmatrix}1&0\\0&1\end{bmatrix} \end{cases} \tag{13b}$$

## Quantum genetic programming (qGP)

Basically, an observer will try to maximize the qEV guided by qDT (nest of hierarchy yes/no logic) to "beat" the nature in the game. A qDT can be optimized by the qGP. The purpose of qGP iterative evolution is to find a satisfactory qDT through learning historical data. The learning rule is as follows:

a) If the "cat" is alive ($q_1$) and an observer bets the "cat" is alive ($a_1$), the observer win a game; if an observer bets the "cat" is dead ($a_2$), the observer lose a game.

b) If the "cat" is dead ($q_2$) and an observer bets the "cat" is dead ($a_2$), the observer win a game; if an observer bets the "cat" is alive ($a_1$), the observer lose a game.

An optimization problem mainly includes the selection of evaluation function and the acquisition of optimal solution. The evaluation function of qDT is a fitness function $f_{fitness}$ (15) based on observed value $V_k$ (14), and the optimal solution is obtained through continuous evolution by using selection, crossover, mutation as in (16) and implemented by qGP algorithm. The quantum expected value of the kth bet is as follows:

$$<V_k> = \begin{cases} p_1\omega_1|\langle a_1|q_1\rangle|^2 = p_1\omega_1 r_{11} = p_1\omega_1, & |\phi\rangle_{mental} = |a_1\rangle \text{ and } |\psi\rangle_{atom} = |q_1\rangle \\ p_1\omega_2|\langle a_1|q_2\rangle|^2 = p_1\omega_2 r_{12} = -p_1\omega_2, & |\phi\rangle_{mental} = |a_1\rangle \text{ and } |\psi\rangle_{atom} = |q_2\rangle \\ p_2\omega_1|\langle a_2|q_1\rangle|^2 = p_2\omega_1 r_{21} = -p_2\omega_1, & |\phi\rangle_{mental} = |a_2\rangle \text{ and } |\psi\rangle_{atom} = |q_1\rangle \\ p_2\omega_2|\langle a_2|q_2\rangle|^2 = p_2\omega_2 r_{22} = p_2\omega_2, & |\phi\rangle_{mental} = |a_2\rangle \text{ and } |\psi\rangle_{atom} = |q_2\rangle \end{cases} \quad (14)$$

$$f_{fitness} = \sum_{k=0}^{N} <V_k> \quad (15)$$

$$qDT \xrightarrow{evolution} \underset{qDT \in (F \cup T)}{argmax}(f_{fitness}) \quad (16)$$

**qGP algorithm**[15-18]

*Input*:
- Historical data set $\{d_k = (q_k, x_k), k = 0, \cdots, N\}$.
- Setting
1) Operation set $F = \{+, *, //\}$
2) Data set $T = \{H, X, Y, Z, S, D, T, I\}$, eight basic quantum gates
3) Crossover probability = 60~90%; Mutation probability = 1~10%.

*Initialization*:
- Population: randomly create 100 ~ 500 qDTs.

*Evolution*:
- for i = 0 to n, n = 50~100 generations
a) Calculate fitness for each qDT based on historical data set.
b) According to the quality of fitness:
   i. Selection: selecting parent qDTs.
   ii. Crossover: generate a new offspring using the roulette algorithm based on crossover probability.
   iii. Mutation: randomly modify parent qDT based on mutation probability.

*Output*:
- A qDT of the best fitness.

**Results**

| run | natural state | qDT$_1$ | qDT$_2$ | ⋯ | qDT$_{47}$ | ⋯ | qDT$_{56}$ | ⋯ | qDT$_{100}$ |
|---|---|---|---|---|---|---|---|---|---|
| 1 | 0 | 1 | 0 | ⋯ | 0 | ⋯ | 0 | ⋯ | 0 |
| 2 | 1 | 0 | 0 | ⋯ | 1 | ⋯ | 0 | ⋯ | 1 |
| 3 | 1 | 0 | 1 | ⋯ | 1 | ⋯ | 0 | ⋯ | 0 |
| ⋮ | ⋮ | ⋮ | ⋮ | | ⋮ | | ⋮ | | ⋮ |
| ⋮ | ⋮ | ⋮ | ⋮ | | ⋮ | | ⋮ | | ⋮ |
| 5000 | 0 | 0 | 1 | ⋯ | 1 | ⋯ | 1 | ⋯ | 0 |
| ⋮ | ⋮ | ⋮ | ⋮ | | ⋮ | | ⋮ | | ⋮ |
| ⋮ | ⋮ | ⋮ | ⋮ | | ⋮ | | ⋮ | | ⋮ |
| 9998 | 0 | 0 | 1 | ⋯ | 0 | ⋯ | 0 | ⋯ | 0 |
| 9999 | 0 | 0 | 1 | ⋯ | 0 | ⋯ | 1 | ⋯ | 1 |
| N=10000 | 1 | 1 | 1 | ⋯ | 0 | ⋯ | 1 | ⋯ | 1 |

**Table 3 100 qDTs of an observer are simulated and optimized by qGP.**

Shown as in Table 3, one hundred qDTs of an observer are simulated by computer and all 100 qDTs are optimized by qGP through learning historical records for this observer as his/her experiences stored in his/her memory. This observer will make decision based on these past experiences (100 qDTs). The average of winning rate is 51.7 and standard derivation of winning rate is 0.2 (Figure 4). The winning rates for all qDTs are almost the same that means the experiences of this observer are very stable.

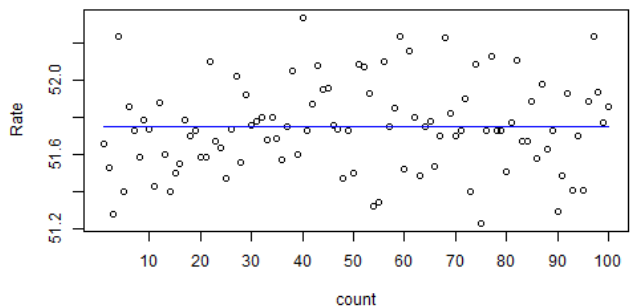

**Figure 4 Winning rates and average for the observer.**

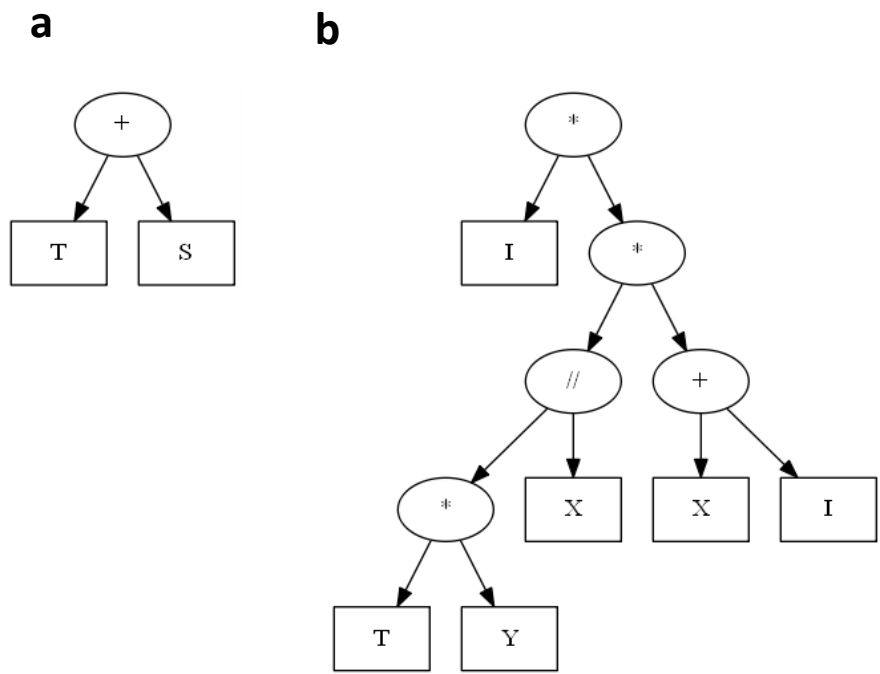

**Figure 5 (a)** qDT$_{47}$ with the smallest expected value **(b)** qDT$_{56}$ with the biggest expected value.

Of 100 qDTs, qDT$_{47}$ got the smallest expected value shown as Figure 5 (a); qDT$_{56}$ got the biggest expected value shown as Figure 5 (b). Based on the qDT$_{47}$ as in (17) (Figure 5 (a)), there are only one simple strategy ($S_1$) that the observer can take, which is 54% degrees of belief to believe the cat is alive and 46% degrees of belief to believe the cat is dead (Figure 6 (a)). We can see this qDT's subjective probability (degrees of belief) close to 50/50 like randomly throw a coin to decide if the cat is alive or dead, so almost no valued information obtained, that's the reason qDT$_{47}$ obtained the smallest expected value. Based on the qDT$_{56}$ as in (18) (Figure 5 (b)), there are two strategies ($S_i \in \{S_1, S_2\}$) that the observer can take, one is 100% degree of belief to believe the cat is alive and the other one is 96% degree of belief to believe the cat is dead and 4% degree of belief to believe the cat is alive (Figure 6 (b)). We can see now the qDT$_{56}$'s subjective probability (degrees of belief) close to unity, so almost maximum information obtained, that's the reason observer$_{56}$ got the biggest expected value.

$$qDT_{47} = (T + S) \tag{17}$$

- $S_1 = (T + S) \rightarrow \hat{V} = 0.54|a_1\rangle\langle a_1| + 0.46|a_2\rangle\langle a_2|$

$$qDT_{56} = \left(I * \left(((T * Y)//X) * (X + I)\right)\right) \tag{18}$$

- $S_1 = \left(I * (X * (X + I))\right) \rightarrow \hat{V} = |a_1\rangle\langle a_1|$
- $S_2 = \left(I * ((T * Y) * (X + I))\right) \rightarrow \hat{V} = 0.04|a_1\rangle\langle a_1| + 0.96|a_2\rangle\langle a_2|$

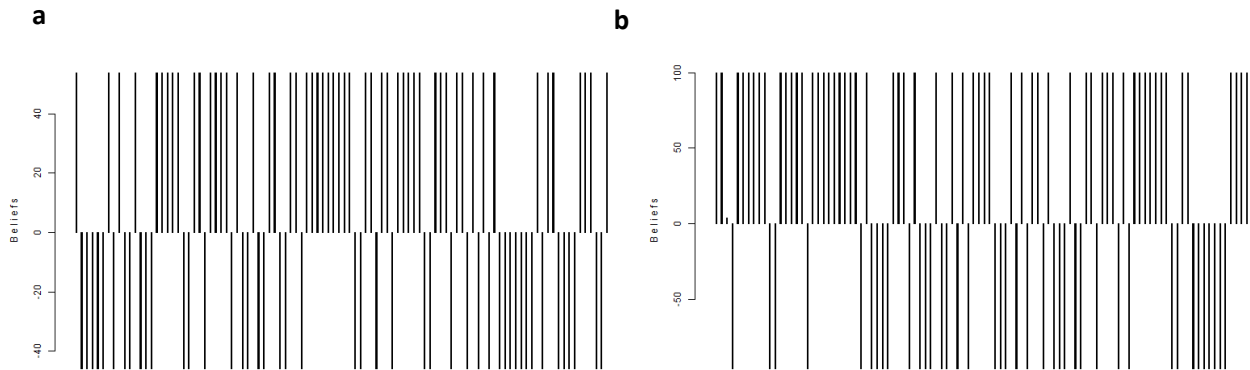

**Figure 6 (a) subjective degrees of belief of** qDT$_{47}$ **(b) subjective degrees of belief of** qDT$_{56}$

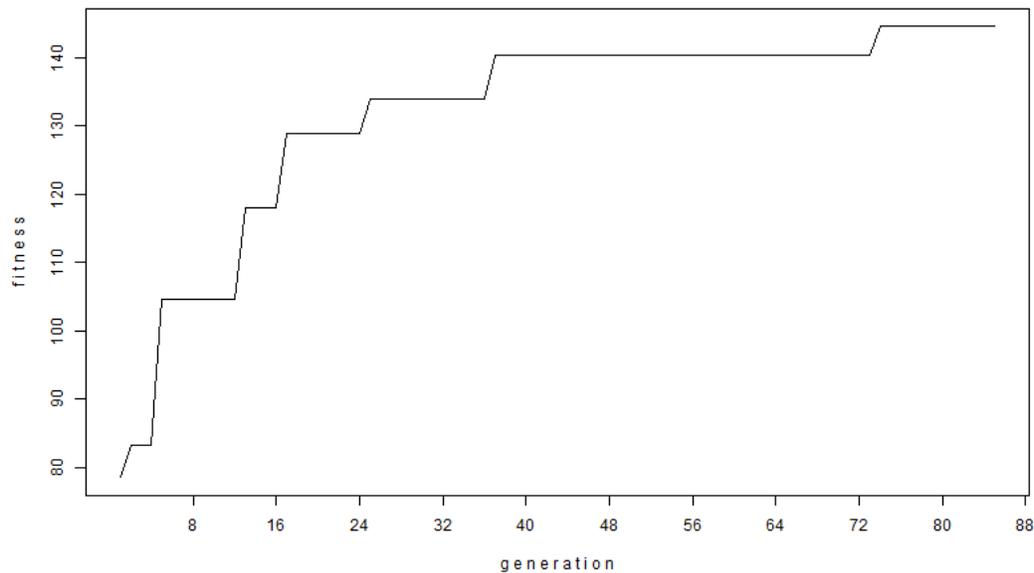

**Figure 7 evolution of** qDT$_{33}$

The expected value for qDT$_{33}$ is between the biggest expected value and smallest expected value. The 88 generations of evolution of the qDT$_{33}$ is shown in Figure 7; the nest of hierarchy structure of qDT$_{33}$ is shown in Figure 8. Based on the qDT$_{33}$ as in (19) (Figure 8), there are four strategies with different subjective degrees of belief that the observer can take. At any given moment the observer's degrees of belief are unknown, the qDT$_{33}$ which simulates observer's degrees of belief can be interpreted as a mixed strategy with four different strategies $\{S_1, S_2, S_3, S_4\}$ for the observer, and the final decision is made by "quantum projection measurement" which the observer's brain selects an action $a_i \in \{a_1, a_2\}$ with degrees of belief from one of four available strategies $S_i \in \{S_1, S_2, S_3, S_4\}$. (Believe cat is alive denote by $a_1$ and believe cat is dead denote by $a_2$). For this observer, either $S_1$ strategy was selected with 50/50 even probability (no valued information was obtained under total uncertainty), or $\{S_2, S_3, S_4\}$ strategies were selected with almost 100% sure (maximum information was obtained). The strategies selected by qDT$_{33}$ are more likely happened in real world, if we know for sure something will happen, we will take an action; if we have no idea what will happen, best we can do is to toss a coin to "hit" the luck.

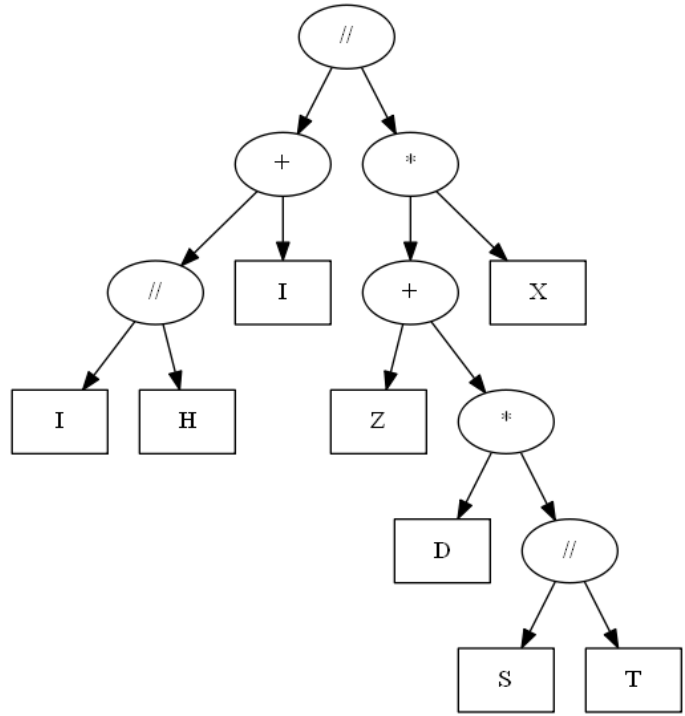

**Figure 8 A nest of hierarchy structure of** qDT$_{33}$.

$$\text{qDT}_{33} = \left(\left((I//H) + I\right)//\left(\left(Z + (D * (S//T))\right) * X\right)\right) \tag{19}$$

- $S_1 = (I + I) \rightarrow \hat{V} = 0.5|a_1\rangle\langle a_1| + 0.5|a_2\rangle\langle a_2|$ (50% belief cat is alive, 50% belief cat is dead)
- $S_2 = (H + I) \rightarrow \hat{V} = |a_1\rangle\langle a_1|$ (100% belief cat is alive)
- $S_3 = \left((Z + (D * S)) * X\right) \rightarrow \hat{V} = 0.05|a_1\rangle\langle a_1| + 0.95|a_2\rangle\langle a_2|$ (5% belief cat is alive, 95% belief cat is dead)
- $S_4 = \left((Z + (D * T)) * X\right) \rightarrow \hat{V} = 0.17|a_1\rangle\langle a_1| + 0.83|a_2\rangle\langle a_2|$ (17% belief cat is alive, 83% belief cat is dead)

The subjective degrees of belief of the first 100 actions the observer took are shown in Figure 9. Detailed information of the first ten actions the observer took is shown in Table 4. For the first action, strategy $S_3$ was applied by the observer who believes that the cat is dead with 95% degrees of belief, the observer got it wrong because the cat is alive when the box is opened; for the fourth action, strategy $S_3$ was applied by the observer who believes that the cat is dead with 95% degrees of belief, this time the observer got it right because the cat is dead when the box is opened; for the eighth action, strategy $S_1$ was applied by the observer who believe that the cat is alive with 50% degrees of belief, the observer got it wrong because the cat is dead when the box is opened; for the tenth action, strategy $S_2$ was applied by the observer who believe that the cat is alive with 100% degrees of belief, the observer got it right this time because the cat is alive when the box is opened.

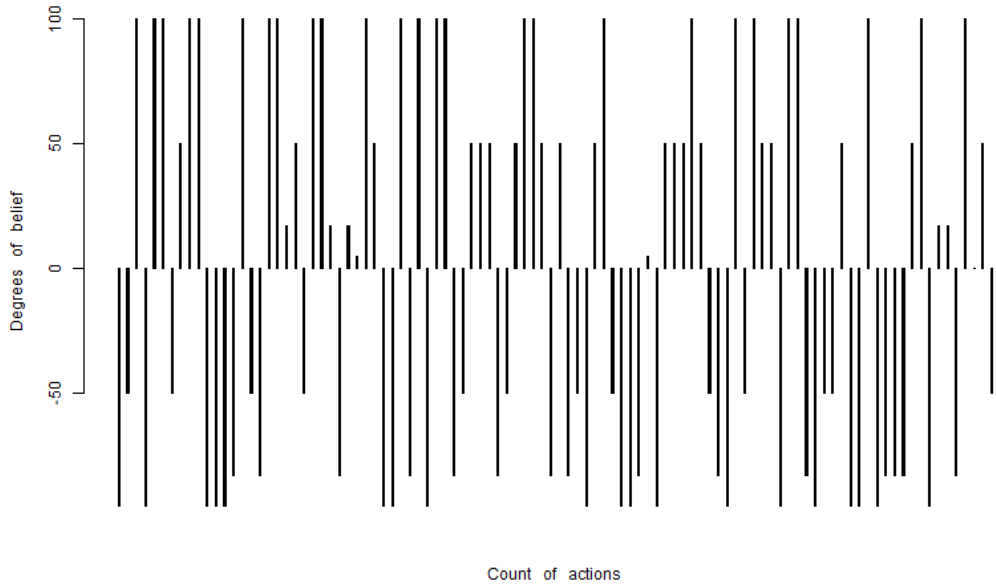

Figure 9 A Observer's degrees of belief ($qDT_{33}$: **positive: believe cat is alive; negative: believe cat is dead**).

| Cat's state | Observer's action selected | Degrees of belief | Strategy | Value |
|---|---|---|---|---|
| $q_1$: alive | $a_2$: believe cat is dead | 95% | $S_3$ | -1 |
| $q_2$: dead | $a_2$: believe cat is dead | 50% | $S_1$ | 1 |
| $q_2$: dead | $a_1$: believe cat is alive | 100% | $S_2$ | -1 |
| $q_2$: dead | $a_2$: believe cat is dead | 95% | $S_3$ | 1 |
| $q_2$: dead | $a_1$: believe cat is alive | 100% | $S_2$ | -1 |
| $q_1$: alive | $a_1$: believe cat is alive | 100% | $S_2$ | 1 |
| $q_2$: dead | $a_2$: believe cat is dead | 50% | $S_1$ | 1 |
| $q_2$: dead | $a_1$: believe cat is alive | 50% | $S_1$ | -1 |
| $q_1$: alive | $a_1$: believe cat is alive | 100% | $S_2$ | 1 |
| $q_1$: alive | $a_1$: believe cat is alive | 100% | $S_2$ | 1 |

Table 4 The first ten actions by $qDT_{33}$.

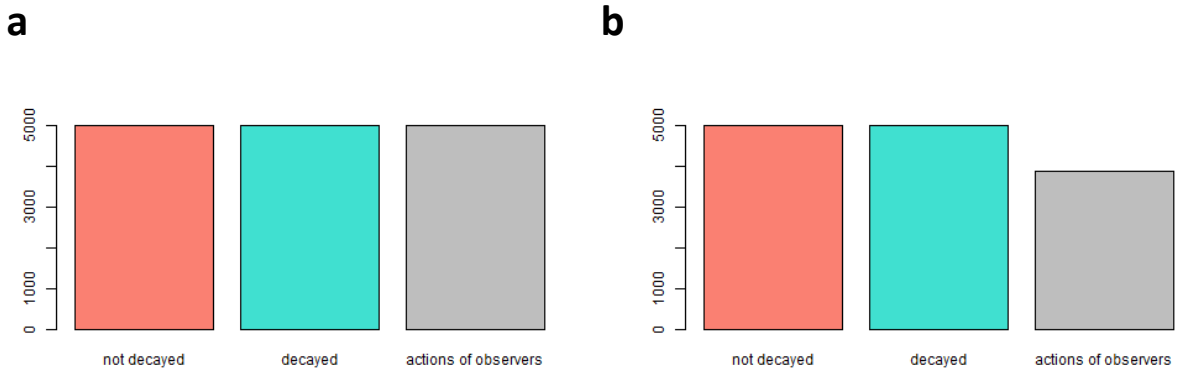

Figure 10 (a) winning rate of 100 optimized qDTs (b) winning rate of 100 qDTs without optimized.

According to Table 1, we can apply the majority rule to let the observer determine the cat's state based on the experience obtained from 100 qDTs; the rules are as follows:

1. If more than 50 qDTs decide cats are alive, the observer believes that the cat is alive;

2. If more than 50 qDTs decide cats are dead, the observer believes that the cat is dead;

Figure 10 (a) shows that the winning rate of the observer jointly determined by the 100 qDTs (the experiences of the observers) optimized based on maximum value is 50%, which is almost completely conforms to the objective frequency of the cat's state (Born rule); however, the winning rate of the observer jointly determined by 100 qDTs (generated by simple coin toss) that are not optimized based on the maximum value is only 39% (Figure 10 (b)), not conforms to the Born rule. The reason is simple: the useful information obtained from the decision of completely random coin toss (50/ 50) is 0, while the jointed decision of 100 qDTs optimized based on maximum value almost obtains the maximum information of the nature, so it conforms to the Born rule; in other words, the observer "finds" the Born rule.

## Discussion

$$\text{entity} = \{\text{state}_i, \text{value}_i, i = 1,2,\cdots N\} \tag{20}$$

Both classical particles and quantum entities can be represented by (20); the current dynamic state of an entity is denoted by $\text{state}_i$, while the corresponding kinematic trajectory is denoted by $\{\text{value}_i\}$. Classical mechanics describes the laws of particles based on the dynamics equations (differential) and their kinematic trajectories (integral). A dynamic equation is essentially a set of logical rules, and the trajectory (historical data) of an entity describes the "historical events" that happened in the past of the entity. By using calculus, classical mechanics fully describe the past and future of entities as successive "events" (certain). The evolution of the dynamic state of a quantum entity is described by the Schrödinger differential equation, but the kinematic trajectories of quantum entities cannot be obtained by integration, and the "position" of a quantum entity can only be obtained by an irreversible "quantum jump" caused by quantum measurements. Quantum mechanics describes past "events" (discontinued) of quantum entities through differential equation and quantum measurements, while future "events" of quantum entities are statistically determined by Born's rule (uncertain).

We can look at quantum measurement in this way: nature asks questions and observers answer natural questions; basically it is a game between nature and observers: there is a sequence of "choices" made by nature, and observers select a sequence of actions guided by optimized strategies to decode the nature.

$$\{d_i, i = 1,2,\cdots N\}: \text{questions posed by Nature.} \tag{21}$$

$$\{qDT_i, i = 1,2,\cdots N\}: \text{observers answer natural questions using yes/no logic in qDTs.} \tag{22}$$

Let go the traditional differential equations dynamic variable approach; by learning the historical "events" given by nature, the yes/no logic of qDT (nest of hierarchy) is applied to evolve the natural dynamics rules, and to interpret the kinematic trajectories of nature as follows:

$$\text{historical events (quantum entity)} \xleftarrow{\text{learn}} \text{computational simulation (observer)} \begin{cases} \xrightarrow{\text{evolve}} \text{dynamics rules} \\ \xrightarrow{\text{interpret}} \text{kinematic trajectories} \end{cases}$$

We divide N=10000 data points ("cat" states) into subset M=100 groups; each subset data group $M_i$ includes L=100 data points, and satisfy:

1. subset $M \ll N$: The 100 data points in subset M are small enough compared to all the data points in N.

2. $M_i = \{L_1, L_2, \cdots, L_{100}\}$: The subset M group data should be large enough to ensure that the objective frequency of the subset M group data is approximately equal to the objective frequency of all the data, so the subset M can approximately represent the overall sample N=10000.

Now the question is: whether we can accurately "reconstruct" the "trajectory" of a quantum entity by learning from all subsets $\{M_i, i = 1,2,\cdots,100\}$?

Just as classical mechanics uses the principle of least action to obtain the trajectories of particles, we use the principle of maximum expected value to approximately obtain the "trajectories" of quantum entities. By maximizing the expected value of the observer, qGP iteratively evolves to obtain a sequence of qDTs each with different strategies in (23).

$$\text{qGP} \xrightarrow{\text{evolove}} \{\text{qDT}_i(s_1, \cdots s_m), i = 1,2, \cdots 100\} \quad (23a)$$

$$\text{qDT}_i = \max_{\text{fitness}} \left( \sum_i \text{qEV}_i / (y_i - x_i) \right), \text{feedback} = (y_i - x_i) \quad (23b)$$

Where $y_i$ denotes observer's subjective value (computed value), $x_i$ denotes nature's objective value (measured value); feedback is the difference between the value computed by the observer and the value measured of nature. The observer adjusts the current value based on past experience while maximizing qEV; qEV is an assessment of the future and feedback is an assessment of the past.

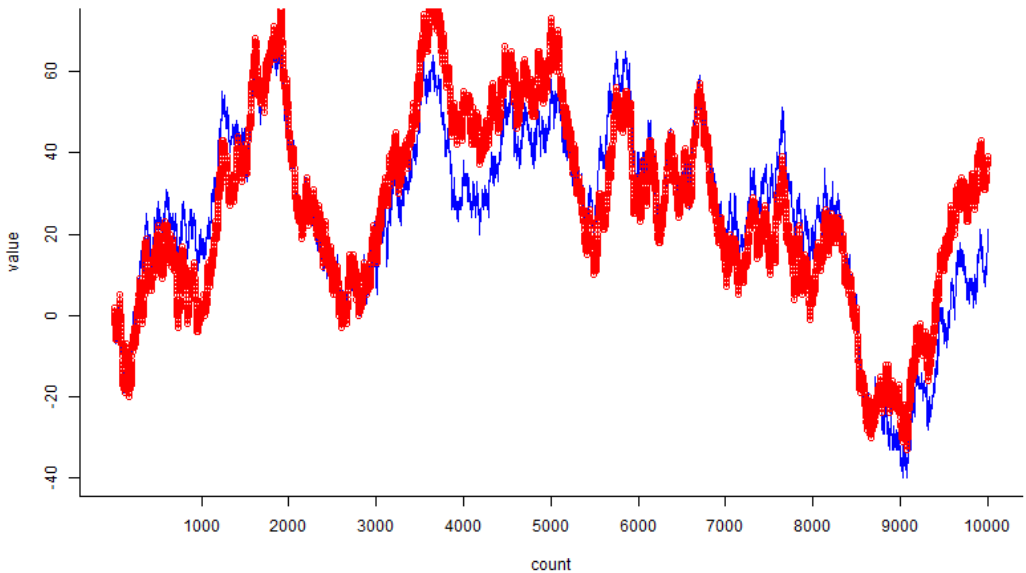

**Figure 11 the "trajectories" of quantum entity reconstructed by** $\{\text{qDT}_i(s_1, \cdots s_m), i = 1,2, \cdots 100\}$.

By answering natural questions, qDTs can "reconstruct" the "trajectories" of quantum entities very well. As shown in Figure 11, we were able to "reconstruct" the "trajectories" of quantum entity with 70% accuracy (blue: nature; red: observer); because we cannot get the "prior" information of quantum entity, the information of quantum entity is incomplete, that is, quantum entity may have an infinite number of " trajectories ", so we cannot accurately predict the future "trajectory" of the quantum entity; It seems that nature is indeed playing games with us, and it is impossible to accurately predict the future "trajectory" of quantum entities unless we can "dance" with nature, can we?